\documentclass[a4paper,11pt]{article}
\pdfoutput=1
\usepackage[utf8x]{inputenc}
\usepackage{jheppub,color,subfigure,slashed,hyperref}
\usepackage{multirow}

\hypersetup{unicode=true,pdftoolbar=true,pdfmenubar=true,pdffitwindow=false,pdfstartview={FitH}, pdfsubject={},pdfcreator={},pdfproducer={}, pdfkeywords={},pdfnewwindow=true,colorlinks=true,linkcolor=blue,citecolor=magenta,filecolor=magenta,urlcolor=cyan}

\newcommand{\be}{\begin{equation}}
\newcommand{\ee}{\end{equation}}
\newcommand{\bea}{\begin{eqnarray}}
\newcommand{\eea}{\end{eqnarray}}
\newcommand{\eq}[1]{Eq.~(\ref{#1})}

\newcommand{\nn}{\nonumber}

\newcommand{\tev}{\text{TeV}}
\newcommand{\gev}{\text{GeV}}
\newcommand{\mev}{\text{MeV}}

\def\stw{s_{\theta_W}}
\def\ctw{c_{\theta_W}}
\def\ttw{t_{\theta_W}}
\def\lra#1{\overset{\text{\scriptsize$\leftrightarrow$}}{#1}}

\makeatletter
\def\section{\@startsection {section}{1}{\z@}{-3.5ex plus -1ex minus
 -.2ex}{2.3ex plus .2ex}{\large\bf}}
\def\subsection{\@startsection{subsection}{2}{\z@}{-3.25ex plus -1ex
minus -.2ex}{1.5ex plus .2ex}{\normalsize\bf}}
\makeatother
\makeatletter

\@addtoreset{equation}{section}

\makeatother

\begin{document}

\begin{flushright} CERN-TH-2021-142, IPPP/21/35 \end{flushright}

\title{High energy lepton colliders as the ultimate Higgs microscopes}
\author[a]{Shankha Banerjee,}
\author[b]{Rick~S.~Gupta,}
\author[b]{Oscar Ochoa-Valeriano,}
\author[b]{Michael~Spannowsky}
\affiliation[a]{CERN, Theoretical Physics Department, CH-1211 Geneva 23, Switzerland,}
\affiliation[b]{Institute for Particle Physics Phenomenology,\\Durham University, South Road, Durham, DH1 3LE, United Kingdom}

\emailAdd{shankha.banerjee@cern.ch}
\emailAdd{sandeepan.gupta@durham.ac.uk}
\emailAdd{oscar.ochoa-valeriano@durham.ac.uk}
\emailAdd{michael.spannowsky@durham.ac.uk}
\date{\today}

\abstract{We study standard electroweak/Higgs processes at the high-energy lepton colliders ILC and CLIC. We identify a subset of three operators in the SMEFT that give leading contributions to these processes at high energies. We then perform a `high-energy fit' including these operators. Our final bounds surpass existing LEP bounds and HL-LHC projections by orders of magnitude. Furthermore, we find that these colliders can probe scales up to tens of TeV, corresponding to the highest scales explored in electroweak/Higgs physics.}

\maketitle

\section{Introduction}
\label{sec:_Introduction}

The main goal of high energy physics is to probe the smallest possible length scales or equivalently the highest accessible energies. In this regard, indirect measurements can often explore much higher energies than direct measurements. A prime example is LEP, which could probe energy scales far beyond its centre-of-mass energy, \textit{i.e.}, up to a few \tev, through its precise measurements at the $Z$-pole. Some of these LEP measurements provide the most powerful bounds on the scale of new physics even today. This is because indirect effects are sensitive to irrelevant operators of the Standard Model Effective Field Theory (SMEFT) (for a non-exhaustive list of SMEFT studies, see Refs.~\cite{Buchmuller:1985jz, Giudice:2007fh, Grzadkowski:2010es, Gupta:2011be, Gupta:2012mi, Banerjee:2012xc, Gupta:2012fy, Banerjee:2013apa, Gupta:2013zza, Elias-Miro:2013eta, Contino:2013kra, Falkowski:2014tna, Englert:2014cva, Gupta:2014rxa, Amar:2014fpa, Buschmann:2014sia, Craig:2014una, Ellis:2014dva, Ellis:2014jta, Banerjee:2015bla, Englert:2015hrx, Ghosh:2015gpa, Degrande:2016dqg, Cohen:2016bsd, Ge:2016zro, Contino:2016jqw, Biekotter:2016ecg, deBlas:2016ojx, Denizli:2017pyu, Barklow:2017suo, Brivio:2017vri, Barklow:2017awn, Khanpour:2017cfq, Englert:2017aqb, panico, Franceschini:2017xkh, Banerjee1, Grojean:2018dqj,Biekotter:2018rhp, Goncalves:2018ptp,Gomez-Ambrosio:2018pnl, Freitas:2019hbk, Banerjee:2019pks, Banerjee:2019twi, Biekotter:2020flu, Araz:2020zyh, Ellis:2020unq, Banerjee:2020vtm, Almeida:2021asy, Chatterjee:2021nms}) that carry the imprint of new physics at energy scales higher than that of the studied process.

Indirect effects due to higher dimensional SMEFT operators scale either as $m_W^2/\Lambda^2$ or $s/\Lambda^2$, $s$ being the centre of mass energy and $\Lambda$ being the scale of new physics. As far as the latter effects are concerned, future lepton colliders such as CLIC and ILC (or in some cases even the LHC, see~\cite{Franceschini:2017xkh, Banerjee1, Araz:2020zyh}) have a clear advantage over LEP given their higher centre-of-mass energy. Relative to low energy measurements at LEP, the same EFT effects would be enhanced by a factor of $s/s_{\mathrm{LEP}}$ (see also  Ref.~\cite{deBlas:2018mhx, Buttazzo:2018qqp, Buttazzo:2020uzc}). As we will show in this work, this will allow lepton colliders to probe scales up to tens of \tev, the highest energy (and smallest length) scale probed in the electroweak/Higgs sectors.

In this work, we will identify the SMEFT operators that give the leading contributions at high energies to the standard electroweak and Higgs processes at lepton colliders like $e^+e^- \to Zh, W^+W^-$ and vector boson fusion (\emph{VBF}) production of the Higgs boson. These will be the EFT effects that are sensitive to the highest possible energy scale. We will assume that the same scale suppresses all irrelevant operators, which will allow us to perform a `high-energy fit', taking into account only this subset of operators. We will see that only three linear combinations of operator coefficients (for a single fermion generation), the so-called `leptonic high energy primaries', give leading contributions to all these processes at high energies. As we will see, the high energy amplitudes for these processes are closely related due to theoretical principles, namely the Goldstone Boson Equivalence theorem that relates the $e^+e^- \to Zh$ and $e^+e^- \to W^+W^-$ processes and the crossing symmetry that connects the \emph{VBF} processes to $e^+e^- \to Zh$.

\section{The leptonic high-energy primaries}
\label{sec:eft}

\begin{table}[t]
\centering
\begin{tabular}{c|cc}
Process & ILC$_{1000}$ & CLIC$_{3000}$\\ \hline \hline \hline
$e^+e^- \to Z\left(\ell^+\ell^-\right)h(\mathrm{all})$ & \checkmark   & \checkmark    \\
$e^+e^- \to e^+e^- h(\mathrm{all})$ & \checkmark & \checkmark    \\
$e^+e^- \to W(2j)W(2j)$ & $\times$ & \checkmark    \\
$e^+e^- \to W(2j)W(\ell\nu_{\ell})$ & $\times$ & \checkmark   
\end{tabular}
\caption{List of processes included in the high-energy fit performed in this work for ILC$_{1000}$ and CLIC$_{3000}$.}
\label{tab:process}
\end{table}

In this section, we will identify the subset of SMEFT operators that give leading contributions to standard electroweak processes at high energy lepton colliders. We will then assume that other SMEFT operators are generated at the same scale and with Wilson coefficients not larger than those of the operators giving dominant high energy contributions. We can thus consider only this subset of operators and perform a `high-energy leptonic fit'. We obtain projections for the case of ILC and CLIC with CoM energy $\sqrt{s}=1$~\tev \, and $\sqrt{s}=3$~\tev, respectively. We also verify numerically that the effect of the other operators is indeed negligible. Following Ref.~\cite{Franceschini:2017xkh}, we call the linear combinations of operator coefficients that enter these high energy amplitudes as the leptonic high-energy primaries.

The processes that we include are $e^+e^- \to Zh$ and the $Z$-boson fusion production (\emph{ZBF}) of Higgs bosons for our ILC projections. For the CLIC projections, we also include $e^+e^- \to W^+W^-$ process.  While the results for the \emph{Zh} and \emph{ZBF} processes are based on our own collider analyses, we use the results of Ref.~\cite{deBlas:2018mhx} for the $e^+e^- \to W^+W^-$ process. We summarise this list of processes including the decay channels we have considered in Table~\ref{tab:process}. 

A remarkable fact about the SMEFT contributions to these processes is that the same set of 3 leptonic high-energy primaries dominate the SMEFT contribution at high energies in each case~\footnote{See, also Ref.~\cite{deBlas:2018mhx} where the Drell-Yan process (not included in this work) has been shown to be sensitive to very high scales. The high-energy amplitude for this process involves an independent set of operators from the ones considered here.}. We thus statistically combine the results from each process to obtain the best bounds on this three-dimensional space \footnote{The EFT corrections to $W$-boson fusion (\emph{WBF}) process $e^+e^- \to \nu_e\overline{\nu}_eh$ at high energies can also be completely described by the same three operators and this process must be included to obtain the best possible bounds. We will keep an analysis of this process for future work.}. We will now describe the above processes in the high-energy limit in SMEFT and discuss the theoretical principles underlying the relationship between them.

\begin{figure}[tbp]
    \centering
    \includegraphics[scale=0.67]{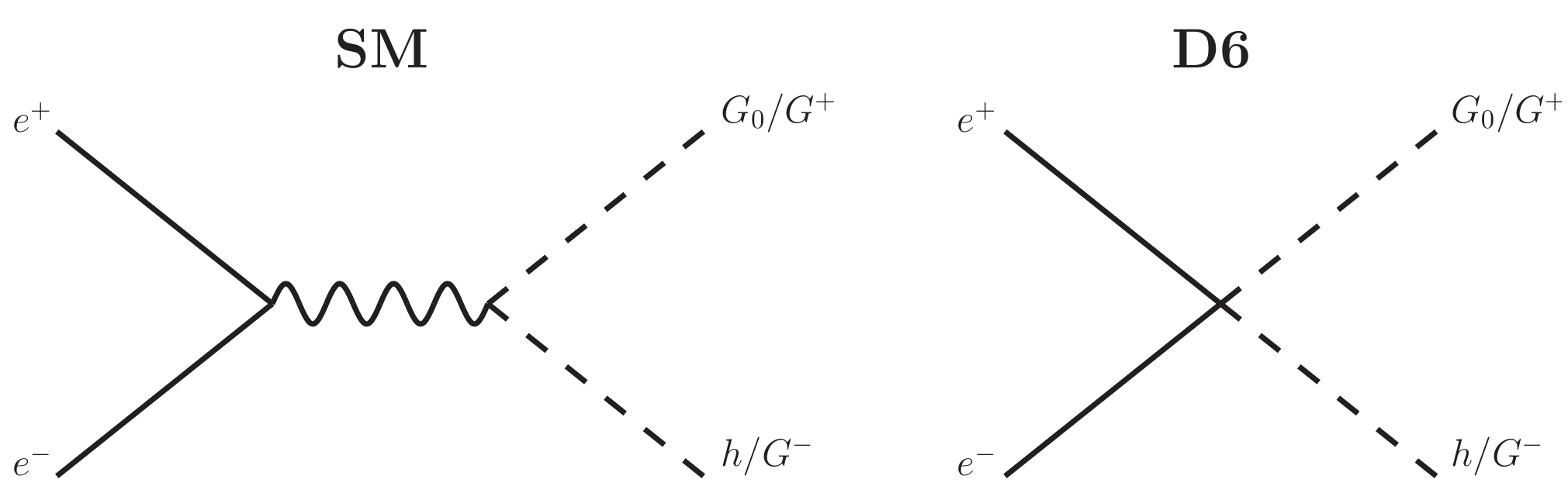}
    \caption{The leading high energy contribution to $e^+e^- \to Zh, W^+W^-$ amplitudes in the SM and D6 SMEFT using the Goldstone Boson Equivalence Theorem.}
    \label{fig:gbe}
\end{figure}

\subsection{High energy \texorpdfstring{$e^+e^- \to Zh, W^+W^-$}{} production in the D6 SMEFT}
\label{sec:_VhWW}

Owing to the Goldstone Boson Equivalence Theorem~\cite{Chanowitz:1985hj, Wulzer:2013mza}, the production of longitudinally-polarised bosons are correlated in the high-energy limit, $E \gg m_W$, where $E=\sqrt{s}$. The externally produced longitudinally-polarised states of the gauge bosons can be represented in the Feynman diagrams by the corresponding Goldstone bosons, up to a factor of $m_W/E$. To understand the high-energy behaviour of these processes, it is, therefore, sufficient to study these in the unbroken phase, where the SM symmetry of $SU(2)_L \times U(1)_Y$ is intact. In the unbroken phase, the electroweak bosons are massless, and the Goldstone bosons reside in the same doublet as the Higgs boson. Hence, it follows that at high energies, the processes, $e^+e^- \to W^+W^-, Zh$--which get their dominant contribution from the longitudinal final-states, $W_LW _L,  Z_L h$--are correlated. This observation was made in the context of the corresponding processes for hadronic colliders in~\cite{Franceschini:2017xkh} (see also Ref.~\cite{Banerjee1}).
 
In the SM, as well as the D6 SMEFT, the high-energy amplitude for these processes can be computed by evaluating the first diagram in Fig.~\ref{fig:gbe} where the dashed lines represent the respective Goldstone bosons. The leading contribution to these processes in the D6 SMEFT, shown in the second diagram in Fig.~\ref{fig:gbe}, arises from $ee hG_0$ and $ee G^+G^-$ contact terms, $G_0, G^\pm$ being the SM Goldstone modes. These contact terms are generated upon expanding the operators in Table~\ref{tab:operators}. As the EFT diagram has no $Z$-propagator, the corresponding amplitude grows quadratically with energy compared to the  SM. The D6 amplitudes for these processes can thus be simply read off from the coefficient of the $ee hG_0$ and $ee G^+G^-$ vertices generated upon expanding the operators in Table~\ref{tab:operators}. This gives, for $s \gg m_Z^2$,
\bea
&&\frac{\delta{\cal A}_{e_R e_R \to WW}}{{\cal A}^{SM}_{e_R e_R \to WW}} = \frac{\delta{\cal A}_{e_R e_R \to Zh}}{{\cal A}^{SM}_{e_R e_R \to Zh}} =\frac{1}{2 q^Z_{e_R}} \frac{s}{m_Z^2} \alpha_{e_R} \nn\\
&&\frac{\delta{\cal A}_{e_L e_L \to Zh}}{{\cal A}^{SM}_{e_L e_L \to Zh}} =\frac{1}{2 q^Z_{e_L}} \frac{s}{m_Z^2} (\alpha_{L1}+\alpha_{L3})\nn\\
&&\frac{\delta{\cal A}_{e_L e_L \to WW}}{{\cal A}^{SM}_{e_L e_L \to WW}} = \frac{1}{2 q^Z_{e_L}} \frac{s}{m_Z^2}  (\alpha_{L1}-\alpha_{L3}),
\label{eqprimaries}
\eea
where ${\cal A}^{SM}$ and $\delta{\cal A}$ are the SM amplitude and the EFT contribution respectively for the given process and $q^Z_f=(T_{3f}-Q_f \stw^2)$. Here, $\alpha_{e_R},\alpha_{L1}$ and $\alpha_{L3}$ are the leptonic high energy primaries which are the linear combinations of Wilson coefficients of D6 SMEFT operators. The `Strongly Interacting Light Higgs' (SILH) Lagrangian~\cite{Giudice:2007fh} is especially suited to parametrise universal new physics effects. The leptonic high energy primaries for the universal case are expressed in terms of the SILH operators as follows,
\bea
\alpha_{L1}&=&\frac{\alpha_{e_R}}{2}=\frac{m_W^2 \ttw^2}{\Lambda^2}\left(c_B+c_{HB}-c_{2B}\right)\nn\\
\alpha_{L3}&=&-\frac{m_W^2}{\Lambda^2}\left(c_W+c_{HW}-c_{2W}\right).
\label{eqsilh}
\eea
We see that only two of these three directions remain independent  in the universal case.

In the Appendix~\ref{sec:_Section3} we show how the Wilson coefficients above can be written in terms of some universal `pseudo-observables'  that can be independently constrained in other collider processes. This allows us to write, 
\bea
\alpha_{L1}&=&\frac{\alpha_{e_R}}{2}=-\ttw^2\left(\delta \kappa_\gamma-\hat{S}-\delta g^Z_1 \ctw^2+Y\right)\nn\\
\alpha_{L3}&=&\delta g^Z_1 \ctw^2+W.
\label{equni}
\eea

For the general case we use the Warsaw basis~\cite{Grzadkowski:2010es} (see second column of Table~\ref{tab:operators}) and the expressions for the high-energy primaries become especially simple,
\bea
\alpha_{e_R}=-\frac{c^e_R v^2}{\Lambda^2},~~~\alpha_{L1}=-\frac{c_L^{l,(1)}v^2}{\Lambda^2}~~\alpha_{L3}=-\frac{c_L^{l,(3)}v^2}{\Lambda^2}.
\label{eqwarsaw}
\eea
Again, using the expressions in \eq{eq:relationC} in Appendix~\ref{sec:_Section3} we rewrite the above equations in terms of the anomalous couplings that can be independently constrained, 
\bea
\alpha_{L1}&=&\frac{\ctw}{g}(\delta g^Z_{e_L}+\delta g^Z_{\nu_L})+\stw^2 \delta g ^Z_1-\ttw^2 \delta \kappa_\gamma,\\
\alpha_{L3}&=&\frac{\ctw}{g}(\delta g^Z_{e_L}-\delta g^Z_{\nu_L})+\ctw^2 \delta g ^Z_1,\\
\alpha_{e_R}&=&\frac{2 \ctw}{g}\delta g^Z_{e_R}+2 \stw^2 \delta g ^Z_1-2 \ttw^2 \delta \kappa_\gamma.
\label{eq:relation}
\eea

\begin{table}[t]
\begin{center}
\small
\begin{tabular}{l|l}
SILH Basis&Warsaw Basis\\\hline\hline\hline
\rule[-1.2em]{0pt}{3em}$\displaystyle{\cal O}_W=\frac{i}{2}\left( H^\dagger  \tau^a \lra {D^\mu} H \right )D^\nu  W_{\mu \nu}^a$&$\displaystyle{\cal O}^{l,(3)}_L= (\bar{L}\sigma^a\gamma^\mu L)(i H^\dagger \sigma^a\lra D_\mu   H)$\\
\rule[-1.2em]{0pt}{3em}$\displaystyle{\cal O}_B=\left( H^\dagger  \lra {D^\mu} H \right )\partial^\nu  B_{\mu \nu}$&$\displaystyle{\cal O}^{l,(1)}_L=  (\bar{L} \gamma^\mu L)(i H^\dagger \lra D_\mu   H)$\\
\rule[-.6em]{0pt}{1.5em}$\displaystyle{\cal O}_{HW}=i g(D^\mu H)^\dagger\sigma^a(D^\nu H)W^a_{\mu\nu}$&
$\displaystyle{\cal O}^e_R= (\bar{e}_R \gamma^\mu e_R)(i H^\dagger \lra D_\mu   H)$\\
\rule[-.6em]{0pt}{1.5em}$\displaystyle{\cal O}_{HB}=i g'(D^\mu H)^\dagger(D^\nu H)B_{\mu\nu}$&\\
\rule[-1.2em]{0pt}{3em}$\displaystyle{\cal O}_{2W}=-\frac{1}{2}  ( D^\mu  W_{\mu \nu}^a)^2$&\\
\rule{0pt}{1.5em}$\displaystyle{\cal O}_{2B}=-\frac{1}{2}( \partial^\mu  B_{\mu \nu})^2$&\\
\end{tabular}
\caption{Dimension-six operators contributing to $e^+e^- \to Zh, W^+W^-$ and \emph{ZBF} Higgs production processes at high energies.}
\label{tab:operators} 
\end{center}
\end{table}

\subsection{\emph{ZBF} Higgs production in the D6 SMEFT}
\label{sec:_ZBF}

\begin{figure}[tbp]
    \centering
    \includegraphics[scale=0.75]{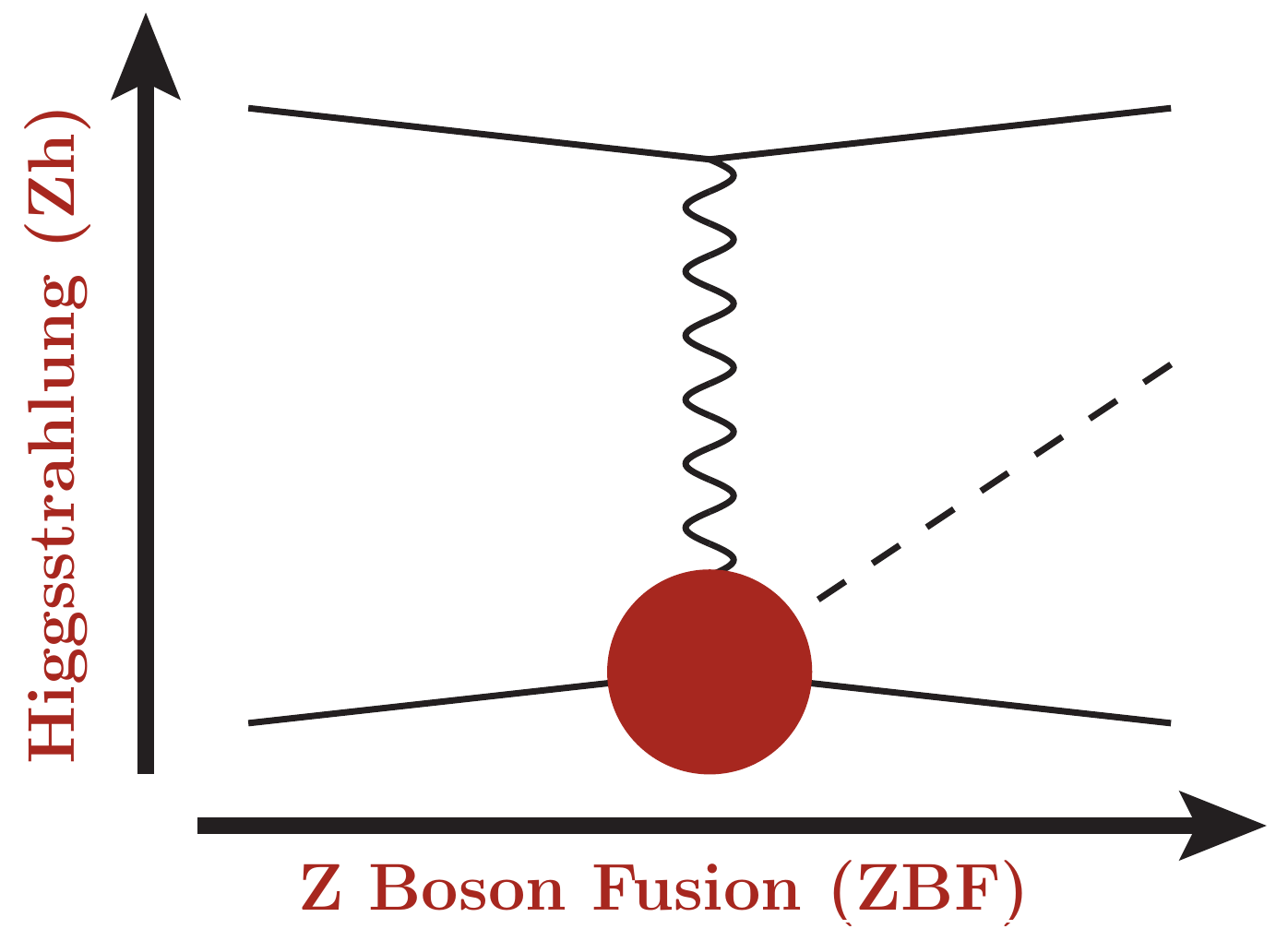}
    \caption{Crossing symmetry that relates the \emph{Zh} and \emph{ZBF} Higgs production processes. The amplitudes of these are the same up to an exchange of the Mandelstam variables $s$ and $t$, $s \leftrightarrow$ t. In consequence, the same direction in SMEFT space control \emph{Zh} at high $s$ and \emph{ZBF} at high $t$.}
    \label{fig:ghZl_operator}
\end{figure}

As far as the \emph{ZBF} process is concerned, the diagram corresponding to it can be obtained just by rotating the one for $e^+e^- \to Zh$ (see Fig.~\ref{fig:ghZl_operator}). In other words, the two processes are related by crossing-symmetry (see also Ref.~\cite{Araz:2020zyh}), and the amplitude for the \emph{ZBF} process can be obtained simply by replacing $s \to t$, \textit{i.e.}, 
\bea
&&\frac{\delta{\cal A}_{e_R Z\to e_R h}}{{\cal A}^{SM}_{e_R Z\to e_R h}} = \frac{\delta{\cal A}_{e_R e_R \to Zh}}{{\cal A}^{SM}_{e_R e_R \to Zh}} (s \to t)= \frac{t}{m_Z^2} \alpha_{e_R} \nn\\
&&\frac{\delta{\cal A}_{e_L Z\to e_L h}}{{\cal A}^{SM}_{e_L Z\to e_L h}} = \frac{\delta{\cal A}_{e_L e_L \to Zh}}{{\cal A}^{SM}_{e_L e_L \to Zh}} (s \to t)= \frac{t}{m_Z^2} (\alpha_{L1}+\alpha_{L3}),
\label{eqprimariesZBF}
\eea
where $s$ and $t$ are the usual Mandelstam variables.

\section{Collider simulation}
\label{sec:_Collider}

We now discuss the collider analysis that helps us in extracting the signal from background processes at future linear $e^+e^-$ colliders~\cite{Yamamoto_2007, Abramowicz:2016zbo, Zarnecki:2020ics}. The single-Higgs production mechanisms we study are $Z\left(\ell^+\ell^-\right)h$, with $\ell = \{e, \mu, \tau\}$, and \textit{ZBF}, targeting a semi-inclusive search of those decays of the Higgs boson that do not yield final-state charged leptons~\footnote{Although it is not an inclusive search, we still keep $\sim 90\%$ of Higgs decays.} to avoid redundancies in the reconstruction of the two-lepton system. Although both processes interfere for final-state electrons, their invariant masses are in themselves powerful discriminating variables that permits us to define mutually exclusive analysis categories. While leptonic decays of the $Z$-boson are suppressed compared to the hadronic ones, we sacrifice event rates for clean final-state signatures. Hence, the signal event topologies are characterised by one pair of opposite-sign same-flavour (OSSF) leptons and additional activity in the form of jets, photons, missing energy, or a combination of the latter.

\subsection{Monte Carlo samples}
\label{subsec:_MonteCarlo}

To obtain bounds on the relevant anomalous couplings, we implement our \texttt{UFO}~\cite{Degrande_2012} model using \texttt{FeynRules}~\cite{Alloul_2014}. We use this model to generate signal and background samples, where we include the interference and the squared terms ensuing from the D6 operators for the signal ones. We consider the EFT-driven $e^+e^- \to \ell^+ \ell^- h$ and $e^+e^- \to e^+ e^- h$ signal processes, where we name the former \textit{Zh}-like and the latter \textit{ZBF}-like. In both cases, we consider the same set of background processes, namely $2\ell\gamma$, $2\ell2\gamma$, $2\ell2\nu_\ell$, $2\ell2\nu_\ell\gamma$, and $2\ell2j$. At high energies, we further take into account $2\ell2V$ ($V=W,Z$), as well as the fully-leptonic $t\bar{t}$ process. For \textit{Zh}-like backgrounds $\ell = \{e, \mu, \tau\}$, while for the \textit{ZBF}-like ones $\ell = \{e, \tau\}$~\footnote{Note that $\nu_{\ell}$ is always the set of all three flavours of neutrinos.}. We neglect reducible backgrounds arising from processes where jets or photons can be misidentified as charged leptons, and consider the SM-driven signal processes as the major sources of background noise.

\begin{table}[t]
\centering
\begin{tabular}{c|ccc}
\hline
Collider & $\sqrt{s}$ [\gev] & $P_{e^+}$, $P_{e^-}$ [\%] & $\mathcal{L}$ $\left[\mathrm{fb}^{-1}\right]$ \\ \hline \hline \hline
ILC$_{250}$ & 250 & $\pm 30, \mp 80$ & 2000  \\
ILC$_{1000}$ & 1000 & $\pm 20, \mp 80$ & 8000 \\
CLIC$_{3000}$ & 3000 & $0, \mp 80$ & 5000
\end{tabular}
\caption{Summary of the main configuration settings for the future linear $e^+e^-$ colliders considered in our study (see text for details). The first column corresponds to the collider, the second column shows the centre-of-mass energy, $\sqrt{s}$, in~\gev, the third column corresponds to the polarisation $P_{e^+}$, $P_{e^-}$ of the $e^+e^-$ beams, and the last column shows the total integrated luminosity, $\mathcal{L}$, in $\mathrm{fb}^{-1}$.}
\label{tab:_SummaryColliders}
\end{table}

\begin{figure}[t]
\centering
\includegraphics[width=1.0\textwidth]{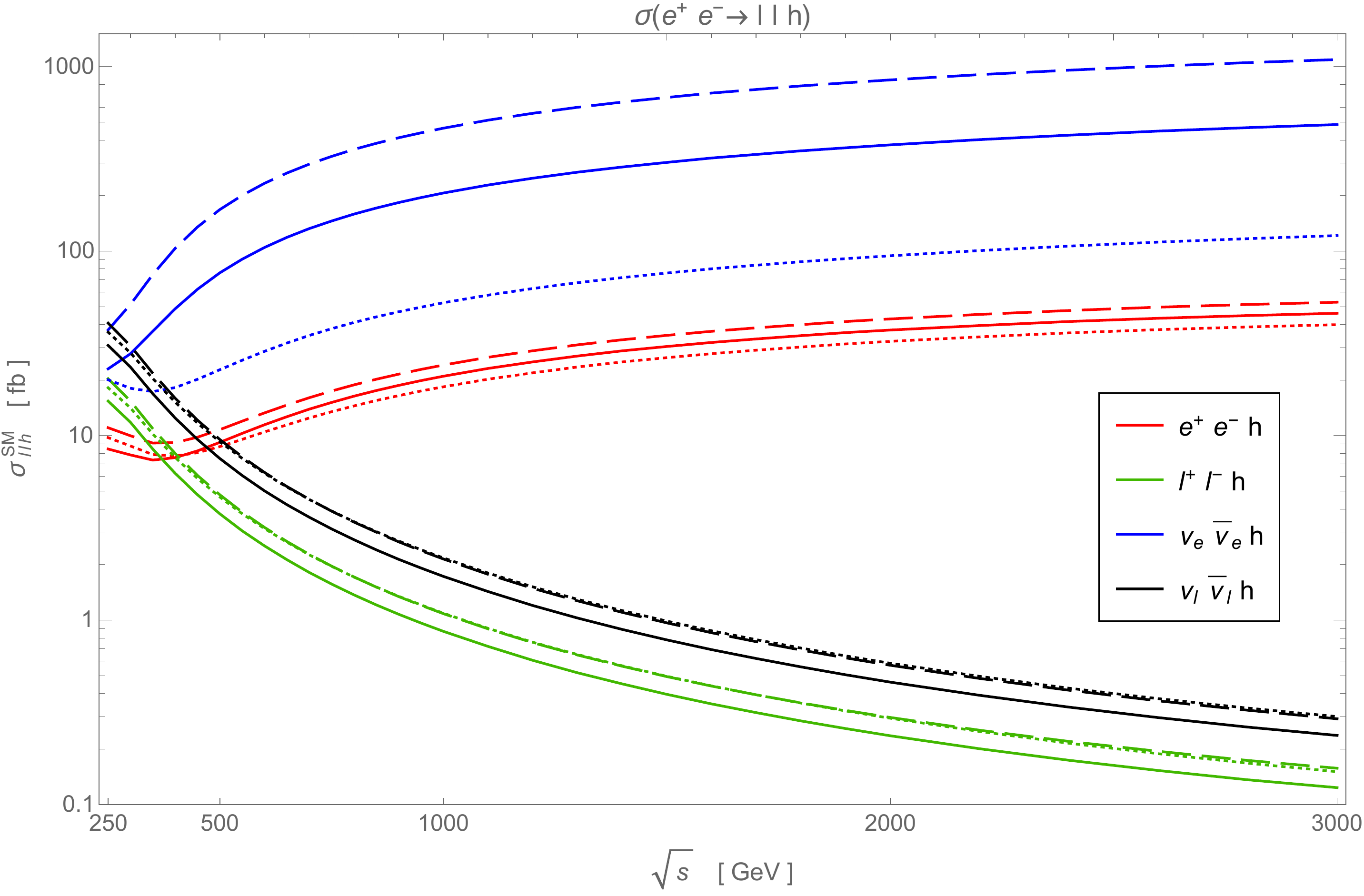}
\caption{Interplay between beam polarisation $P_{e^+}, P_{e^-}$ and cross-sections, $\sigma^{\text{SM}}_{\ell\ell h}$, in fb, as functions of the centre-of-mass energy, $\sqrt{s}$, in \gev, for single-Higgs production in association with a pair of leptons, where in the legend, $\ell = \{\mu, \tau\}$, for unpolarised beams (solid), $P_{e^+}, P_{e^-} = +50\%, -50\%$ (dashed), and $P_{e^+}, P_{e^-} = -50\%, +50\%$ (dotted). Note that both $e^+e^-h$ and $\nu_e\overline{\nu}_eh$ final states consist of \emph{Zh-ZBF} interference for the former, and \emph{Zh-WBF} for the latter. The chosen beam polarisations are meant for illustration purposes and do not necessarily correspond to the projected configuration of future colliders.}
\label{fig:Higgs_XS}
\end{figure}

We consider the centre-of-mass energies $\sqrt{s} = \{250, 1000, 3000\}$~\gev, where the first and second entries correspond to the low- and high-energy phases of the International Linear Collider (ILC)~\cite{Behnke:2013xla, Asner:2013psa}, respectively, and the third one corresponds to the high-energy run of the Compact Linear Collider (CLIC)~\cite{Linssen:1425915, deBlas:2018mhx}. We use, from now on, the nomenclature ILC$_{250, 1000}$ (CLIC$_{3000}$) to refer to the ILC (CLIC) collider run at the $\sqrt{s}$ subscript. We further follow the recommendations of Ref.~\cite{de_Blas_2020} to incorporate longitudinal beam polarisation~\cite{Moortgat-Pick:2005jsx, Durieux:2017rsg, Fujii:2018mli, List:2020wns}~\footnote{Throughout the text, the term \emph{polarisation} stands for longitudinal polarisation, and we omit \emph{longitudinal} for simplicity.} at the considered energies and luminosities to carry out our study. Denoting $\mathcal{L}$ as the integrated luminosity in $\mathrm{fb}^{-1}$, and $P_{e^+}, P_{e^-}$\footnote{In this notation, $+100\%$ corresponds to a fully right-handed polarised beam, and $-100\%$ to a fully left-handed polarised beam.} as the polarisations of the positron and electron beam, respectively, the collider- and $\sqrt{s}$-dependent configurations we take into account are ILC$_{250}$: $\mathcal{L} = 2000, P_{e^+}, P_{e^-} = \pm30\%, \mp80\%$; ILC$_{1000}$: $\mathcal{L} = 8000, P_{e^+}, P_{e^-} = \pm20\%, \mp80\%$; and CLIC$_{3000}$: $\mathcal{L} = 5000, P_{e^+}, P_{e^-} = 0\%, \mp80\%$\footnote{Note that at a given $\sqrt{s}$ we have two polarisation combinations, to each of which corresponds a different fraction of the integrated luminosity, \textit{i.e.}, two colliders in one with different luminosities each.}. We call \textit{left} and \textit{right} polarisation the cases where $P_{e^+} \geq 0, P_{e^-} < 0$ and $P_{e^+} \leq 0, P_{e^-} > 0$, respectively. The interplay between single-Higgs production cross-section $\sigma^{\text{SM}}_{\ell\ell h}$ in association with a pair of leptons and beam polarisation $P_{e^+}, P_{e^-}$ in $e^+e^-$ collisions as a function of the centre-of-mass energy $\sqrt{s}$ is shown in Fig. \ref{fig:Higgs_XS}~\footnote{We impose minimal generator-level cuts to compute these, such as $p_T^{\ell} > 1.0$~\gev, $\left|y^\ell\right| < 5.0$, and $\Delta R_{\ell^+,\ell^-} > 0.1$.}, and a summary of the future colliders considered in this work is shown in Table~\ref{tab:_SummaryColliders}.

To estimate the sensitivity reach of our study at the given energies and luminosities, we use \texttt{MadGraph5\_aMC@NLO}~\cite{Alwall_2014} to generate the leading-order (LO) Monte Carlo event samples, and then shower/hadronise our events with \texttt{Pythia~8.2}~\cite{Sj_strand_2015}. \texttt{DELPHES~3}~\cite{de_Favereau_2014} is used to perform a fast detector simulation with the ILD Tune~\cite{Behnke:2013lya, demin_selvaggi_2016} for the ILC$_{250}$ and ILC$_{1000}$ scenarios, while for CLIC$_{3000}$ we use the CLICdet Stage3 Tune~\cite{Arominski:2649437, Roloff:2649439}. We impose a minimal set of cuts at the generator level on the final-state objects, namely $p_T^{\ell,j,\gamma} > 5$~\gev, $m_{jj} > 10$~\gev, $\left|y^{j,\gamma}\right| \left(\left|y^\ell\right|\right) < 5 \, (3)$, and $\Delta R_{a, b} > 0.1$, where $\Delta R_{a,b} = \sqrt{(\Delta \phi)^2 + (\Delta y)^2}$ is the angular distance in the $\phi-y$ plane\footnote{Although it is customary to use the polar angle $\theta$ in lepton colliders, we use instead the rapidity $y$ for consistency with the generator and detector simulator parameters. In the massless limit $\theta \equiv 2 \arctan(e^{-y})$.} between any two final-state objects $a,b$. We further require a cut on the invariant mass $m_{\ell^+\ell^-}$ of the dilepton system in order to classify our Monte Carlo samples on the mutually exclusive categories \textit{Zh}- and \textit{ZBF}-like-- for the \textit{Zh}-like samples, $m_{\ell^+\ell^-} \in [70, \, 110]$~\gev, whereas for the \textit{ZBF}-like ones we impose a $\sqrt{s}$-dependent cut, $m_{\ell^+\ell^-}^{\sqrt{s}}$, as follows: $m_{\ell^+\ell^-}^{250} > 100$~\gev, $m_{\ell^+\ell^-}^{1000} > 300$~\gev, and $m_{\ell^+\ell^-}^{3000} > 1000$~\gev. An additional cut $y^{\ell^+}\cdot y^{\ell^-} < 0$ is imposed on the \textit{ZBF}-like samples to ensure that the final-state leptons lie in opposite hemispheres of the detector.

The ability of a lepton collider to exploit the four-momentum conservation given that the $p_{e^+e^-}$ initial state is well known allows us to uniquely specify the decay of the Higgs boson,
\begin{equation}
    \label{eq:_FourMomentum}
    p_h = p_{e^+e^-} - p_{\ell^+\ell^-},
\end{equation}
where $p_{\ell^+\ell^-}$ corresponds to the four-momentum of the final-state dilepton system. This allows us to perform model-independent studies of the Higgs width, $\Gamma_h$, and its inclusive production rate, among others. In practice we make use of the dimensionful variable derived from \eq{eq:_FourMomentum} called the \textit{recoil mass}, $m^2_{\text{recoil}}$, as a function of the centre-of-mass energy $\sqrt{s}$, the invariant mass $m_{\ell^+\ell^-}$, and the energy $E_{\ell^+\ell^-}$ of the final-state dilepton system,
\begin{equation}
    \label{eq:_RecoilMass}
    m^2_{\text{recoil}} \equiv s - 2\sqrt{s}E_{\ell^+\ell^-} + m_{\ell^+\ell^-}^2,
\end{equation}
as a discriminator against background noise. It is expected that the $m_{\text{recoil}}$ distribution shows a narrow peak centred at $m_h = 125$~\gev \, given the small width $\Gamma_h \sim 4.088$~\mev \, of the Higgs boson, in line with the \textsc{LHC HXSWG} Report 4~\cite{LHCHiggsCrossSectionWorkingGroup:2016ypw}, provided good detector resolution for the leptons' momenta is in place. Initial- (ISR) and final-state radiation (FSR), as well as beamstrahlung, produce collinear photons that can smear the $m_{\text{recoil}}$ distribution-- experimentally this effect can be controlled by imposing selection cuts on the $p_T$ of the outgoing dilepton system, since photons escaping the detector acceptance will not contribute to a large amount of transverse momentum of the observed objects.

\subsection{\textit{Zh} channel event selection}
\label{subsec:_Zhprod}

We perform a simplified cut-and-count analysis that allows us to suppress background processes from the SM-driven $e^+e^- \to \ell^+\ell^-h$ at ILC$_{250, 1000}$ and CLIC$_{3000}$\footnote{Similar studies focusing on inclusive or $h \to b\bar{b}$ Higgs decays can be found in, \textit{e.g.}, Refs.~\cite{Amar:2014fpa, Craig_2016, Yan:2016xyx}.} on the stable final-state objects after running a detector simulation (see Sec.~\ref{subsec:_MonteCarlo}). The Higgs-strahlung channel is an excellent way to identify on-shell $Z$-bosons and make use of \eq{eq:_FourMomentum} to reconstruct the Higgs boson in a ``model-independent'' way. Being an $s$-channel process, its cross-section decreases with the centre-of-mass energy as $\sigma \sim 1/s$, as previously shown in Fig.~\ref{fig:Higgs_XS}. As the dominant production mode at low energies, the ILC$_{250}$ is a Higgs factory that provides high statistics for the study of gauge-Higgs couplings with leptonically decaying $Z$-bosons that yield a clean signature. However, at high energies, the cross-section is suppressed with respect to the $t$-channel processes, and hadronically decaying $Z$-bosons can enhance the total cross-section since its branching ratio is the largest. Although the hadronic modes of the $Z$-boson dominate over the leptonic ones, $Z \to q\overline{q}$ decays are less clean. As we target those Higgs decays that do not yield charged final-state leptons, QCD activity associated with $Z$ decays might present challenges in reconstructing the latter, requiring detailed background analysis.

\begin{figure}[t]
    \centering
    \includegraphics[scale=.375]{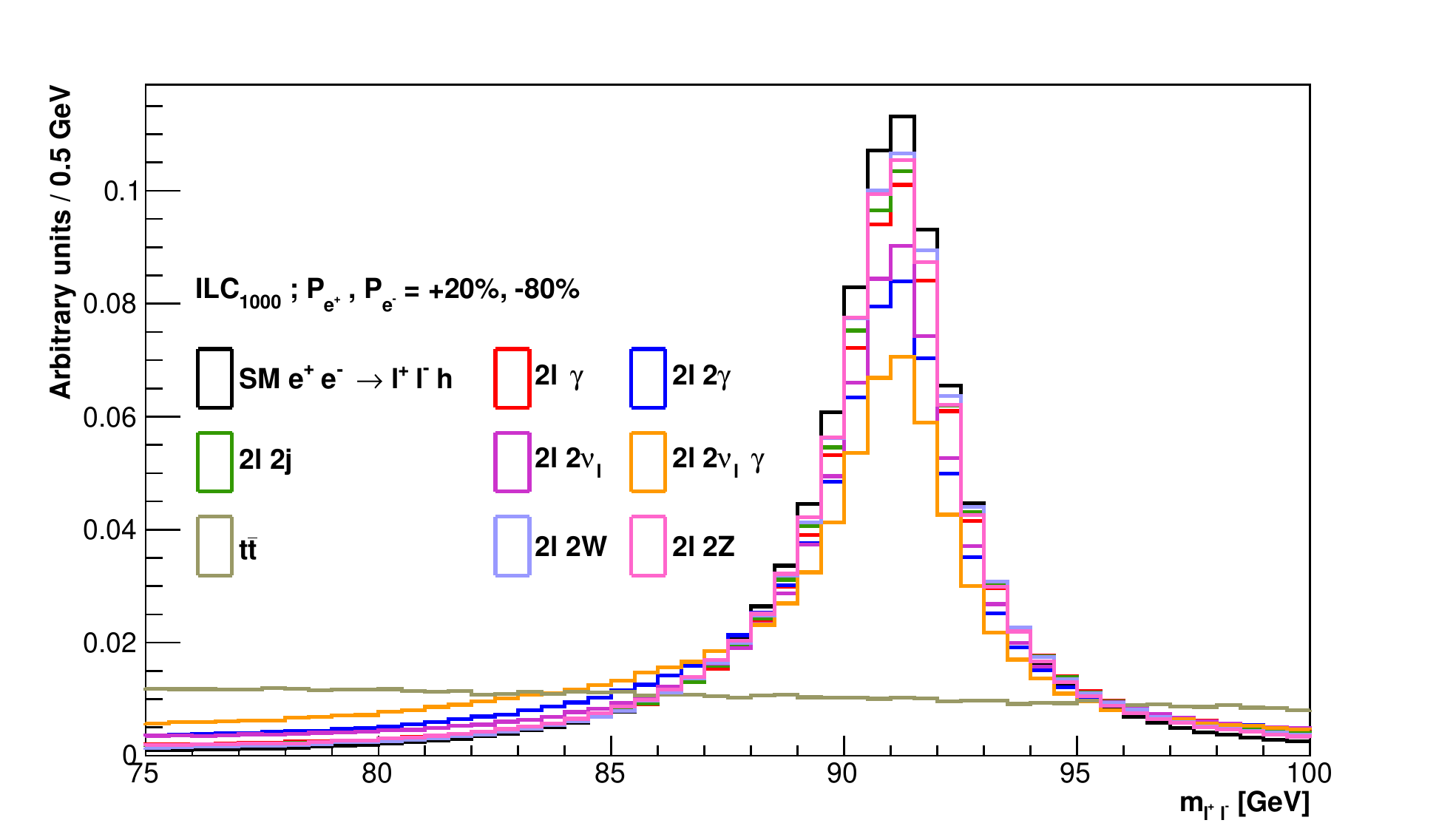}
    \includegraphics[scale=.375]{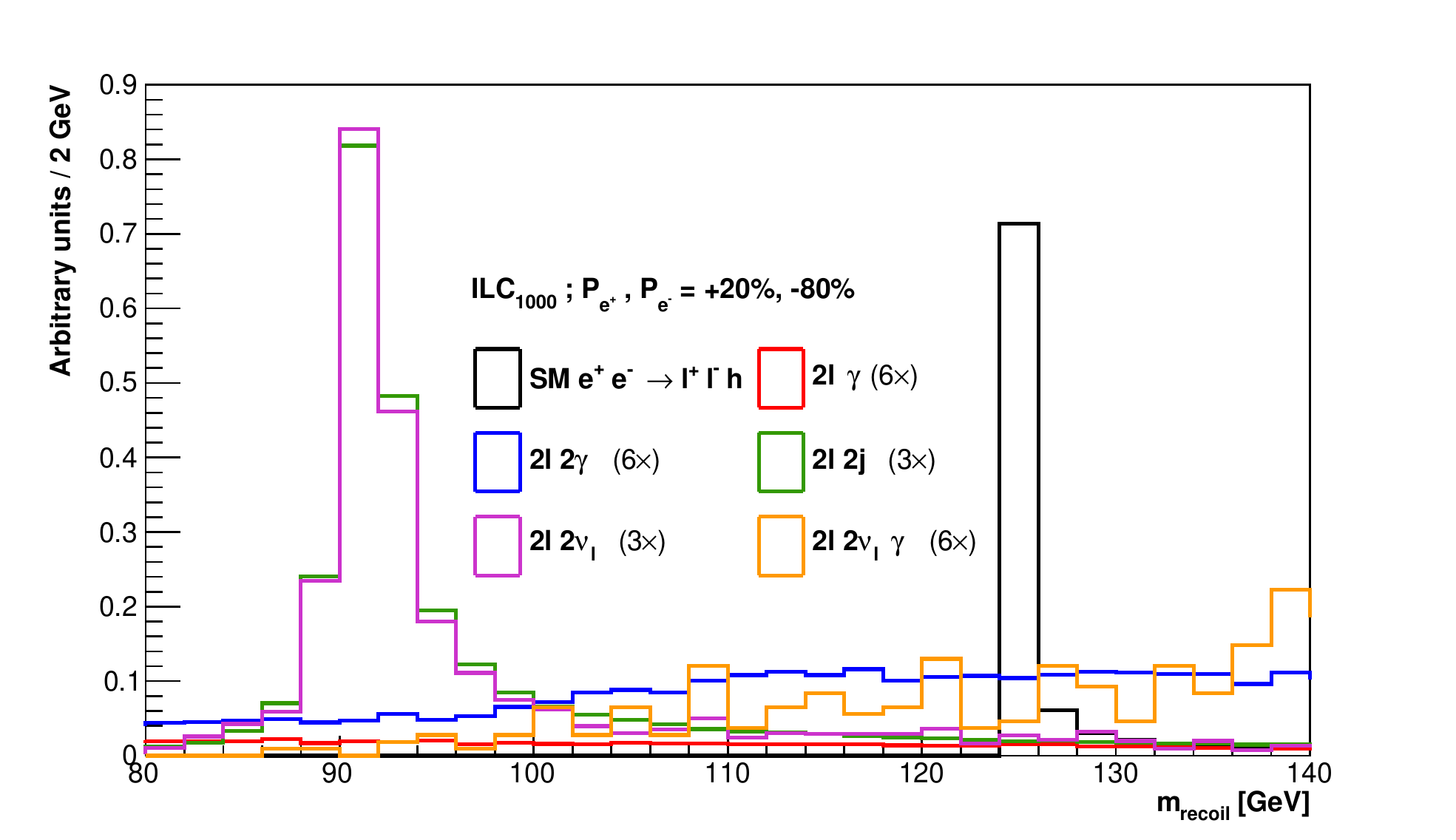}
    \caption{Some plots of the ILC$_{1000}$ \textit{Zh} selection with the \textit{left} polarisation for the SM-driven $e^+e^- \to \ell^+\ell^-h$ (black), $2\ell\gamma$ (red), $2\ell2\gamma$ (blue), $2\ell2j$ (green), $2\ell2\nu_{\ell}$ (magenta), $2\ell2\nu_{\ell} \gamma$ (orange), $t\overline{t}$ (brown), $2\ell2W$ (lilac), and $2\ell2Z$ (pink) processes. All histograms are normalised to unity. \textbf{Left:} $m_{\ell^+\ell^-}$ distribution, peaking at the $Z$-boson mass, before applying the $m_{\ell^+ \ell^-} \in [86.0, 96.0]$~\gev \, cut. \textbf{Right:} $m_{\text{recoil}}$ distribution, with the black curve peaking narrowly at $m_h$ on top of a continuum background, before requiring $m_{\text{recoil}} \in [123.0, 127.0]$~\gev. At this stage of the selection the $t\overline{t}$ background is already negligible, and the distributions of the $2\ell2V$ processes lie beyond the right edge of the plot. Some histograms in the right panel are scaled ($3\times$ or $6\times$) after normalisation for visualisation purposes.}
    \label{fig:Zh_selection}
\end{figure}

A minimal set of selection cuts is applied to the Monte Carlo samples and varies slightly depending on the $\sqrt{s}$ under consideration. We demand the presence of exactly one pair of OSSF leptons with $m_{\ell^+ \ell^-} \in [86, 96]$~\gev-- the resolution achievable at lepton colliders provides a clean reconstruction of the $Z$-boson peak $\sim 91$~\gev, regardless of the collision energy. At the ILC$_{250, 1000}$ and CLIC$_{3000}$ we require the transverse momentum of the dilepton system $p_T^{\ell^+\ell^-} \in \{[40, 70], [400, 490], [1350, 1500]\}$~\gev, respectively. Finally, from \eq{eq:_RecoilMass}, we apply a hard cut on the recoil system's mass $m_{\text{recoil}} \in [123, 127]$~\gev, which makes use of the fact that the $m_{\text{recoil}}$ distribution shows a narrow peak at $m_h \sim 125$~\gev, on top of a continuum background. The $m_{\ell^+\ell^-}$ and $m_{\text{recoil}}$ distributions for the ILC$_{1000}$ with \textit{left} polarisation can be seen in Fig.~\ref{fig:Zh_selection}, and the impact of the event selection on the cross-sections for the dominant SM processes under consideration is shown in Table~\ref{tab:_CutsZh} for the three centre-of-mass energies, as well as for the \textit{left} and \textit{right} polarisations. Since we are interested in contact operators that grow with energy, we find that the ILC$_{250}$ results are not sensitive to the effects of these operators. Hence, we will only focus on the high-energy regime, i.e., ILC$_{1000}$ and CLIC$_{3000}$. It is important to highlight that the kinematic distributions used to perform the event selection show similar shapes/behaviour in both polarisation settings.

\begin{table}[t]
\centering
\resizebox{\columnwidth}{0.9\height}{
\begin{tabular}{|c|c|c|c|c|c|c|}
\hline
$\sigma_{\sqrt{s}}^{\text{stage}}$ [fb] & SM & $2\ell \gamma$ & $2\ell 2\gamma$ & $2\ell 2\nu_{\ell}$ & $2\ell 2\nu_{\ell} \gamma$ & $2\ell 2j$ \\
\hline
\hline
\hline
$\sigma^{\text{in}}_{250}$ & 29.85/26.26 & 5107.28/4735.02 & 316.95/287.56 & 651.88/101.26 & 68.75/8.21 & 264.22/181.23 \\
$\sigma^{\text{out}}_{250}$ & 6.99/6.14 & $< 10^{-6}$ & 0.40/0.32 & 1.02/0.12 & 0.13/0.04 & 0.22/0.08 \\
\hline
$\sigma^{\text{in}}_{1000}$ & 1.45/1.45 & 105.60/87.80 & 19.46/16.48 & 318.14/37.86 & 39.18/4.91 & 28.24/17.80 \\
$\sigma^{\text{out}}_{1000}$ & 0.33/0.31 & 0.017/0.013 & 0.0064/0.0053 & 0.0046/0.0024 & 3/2 $\left(\times 10^{-4}\right)$ & 0.0140/0.0072 \\
\hline
$\sigma^{\text{in}}_{3000}$ & 0.17/0.17 & 6.17/5.09 & 1.65/1.37 & 376.87/43.02 & 61.51/7.18 & 2.29/1.31 \\
$\sigma^{\text{out}}_{3000}$ & 0.026/0.025 & 12/9 $\left(\times 10^{-4}\right)$ & 9.64/6.09 $\left(\times 10^{-5}\right)$ & 3.8/1.9 $\left(\times 10^{-4}\right)$ & 61.5/3.6 $\left(\times 10^{-6}\right)$ & 8.3/4.9 $\left(\times 10^{-4}\right)$\\
\hline
\end{tabular}}
\caption{Cross-sections, $\sigma$, in fb, for the \textit{Zh}-like SM-driven signal $e^+e^- \to \ell^+ \ell^- h$ and its dominant backgrounds before (\textit{in}) and after (\textit{out}) event selection at $\sqrt{s} = \{250, 1000, 3000\}$~\gev. We consider the \textit{left} and \textit{right} beam polarisations $P_{e^+}$,$P_{e^-}$ for each $\sqrt{s}$, reported as \textit{left/right} (see text for details). For the $t\bar{t}$, $\ell^+\ell^-W^+W^-$, and $\ell^+\ell^-ZZ$ channels, $\sigma^{\text{in}} \; \textrm{(in fb)} \sim$  3.47/0.62, 4.44/0.41, and 0.22/0.12, respectively at 1 TeV. At 3 TeV, the respective numbers are 0.08/0.01, 1.77/0.21, and 0.06/0.03. For all cases, $\sigma^{\text{out}} < 10^{-6}$ fb. For $\sqrt{s} = 250$ GeV, the kinematic phase space is not open for any of these three channels.}
\label{tab:_CutsZh}
\end{table}

\subsection{\textit{ZBF} channel event selection}
\label{subsec:_ZBFprod}

We now turn our attention to the $e^+e^- \to e^+e^-h$ \textit{ZBF} mechanism\footnote{On itself, this process seems not to be of much interest in the literature~\cite{GUNION199879, PhysRevD.63.096007, Han:2015ofa}.}, a $t$-channel process with a cross-section that grows with energy as $\sigma \sim \log^2(s/m_Z^2)$. At the ILC$_{250}$ this process interferes constructively with the \textit{Zh} channel, and the reduced phase-space makes it difficult to disentangle the kinematic features, such as a high-energy electron-positron pair in opposite regions of the detector, or $m_{e^+e^-}$ well above $m_Z$. However, from Fig.~\ref{fig:Higgs_XS} it can be seen that at $\sim 500$~\gev \, its cross-section is already larger than the \textit{Zh} one. At the same time, the latter is characterised by an on-shell $Z$-boson (see Fig.~\ref{fig:Zh_selection}). The former features a highly energetic forward/backward electron-positron pair. So it provides access to observables involving the $Z$ and $h$ bosons at high-energy $e^+e^-$ colliders, such as ILC$_{1000}$ and CLIC$_{3000}$, where beam polarisation also plays an important role in enhancing the cross-section of chirality-dependent processes.

A simplified cut-based analysis is performed on the stable final-state objects after running a detector simulation (see Sec.~\ref{subsec:_MonteCarlo}) on the \textit{ZBF}-like Monte Carlo samples at ILC$_{250, 1000}$ and CLIC$_{3000}$. As before, we use $m_{e^+e^-}$ along with $m_{\text{recoil}}$ in \eq{eq:_RecoilMass} as the main background-discriminating variables, since $m_{\text{recoil}}$ displays the Higgs resonance as a sharp, narrow peak on top of a continuum background, and the broad $m_{e^+e^-}$ distribution tends to favour large values well above $m_Z$ at ILC$_{1000}$ and CLIC$_{3000}$ energies, while for ILC$_{250}$ there is large contamination from the \textit{Zh} process. As the centre-of-mass energy, $\sqrt{s}$, becomes larger, the final-state leptons tend to have larger values of $\left|y^e \right|$; however, a major experimental limitation is the absence of electromagnetic calorimeters available for electron tracking and reconstruction in the forward region of the detectors, reducing the phase-space (and in consequence, the cross-section) available for the study and characterisation of this process.

\begin{figure}[t]
    \centering
    \includegraphics[scale=.375]{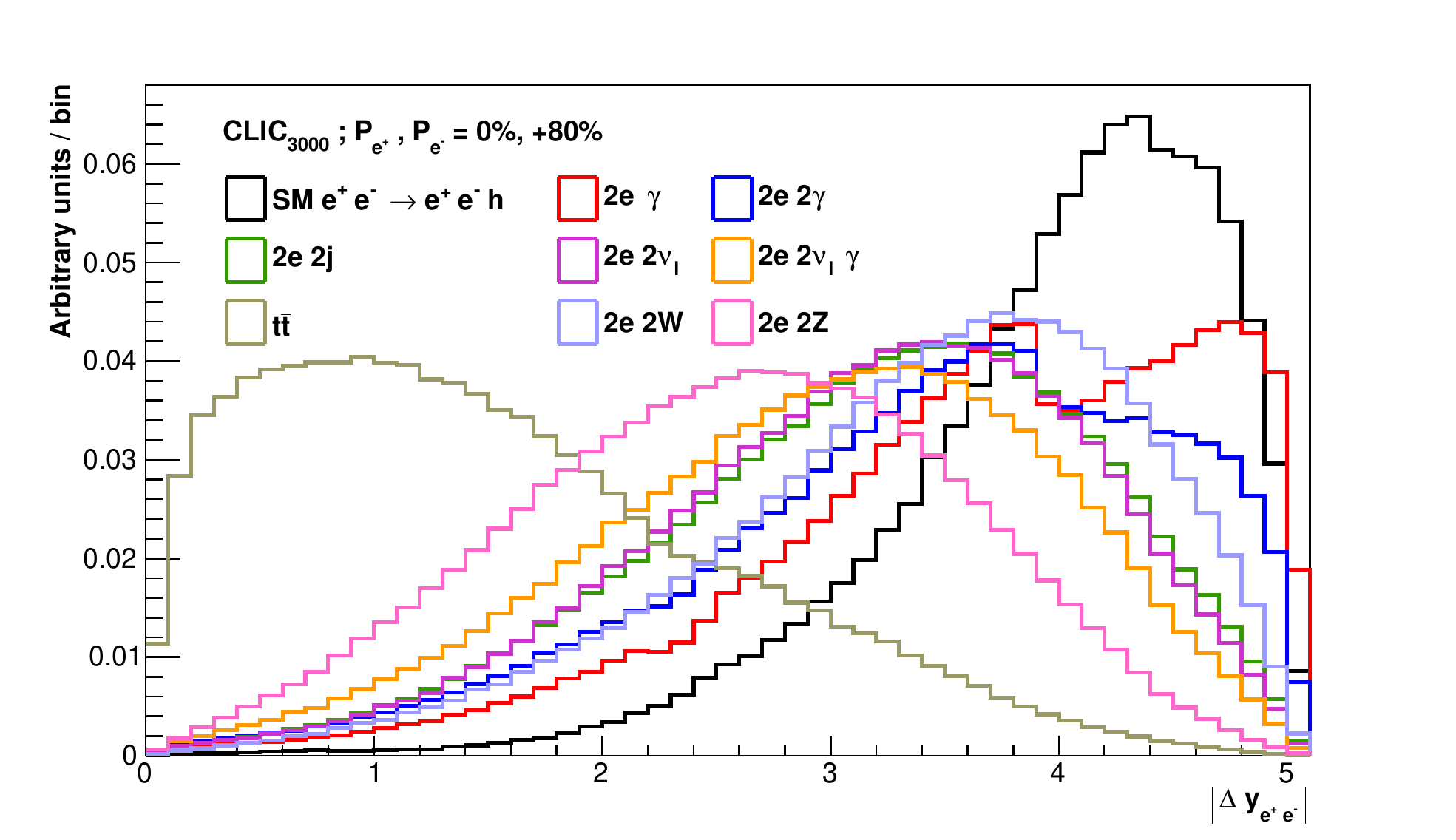}
    \includegraphics[scale=.375]{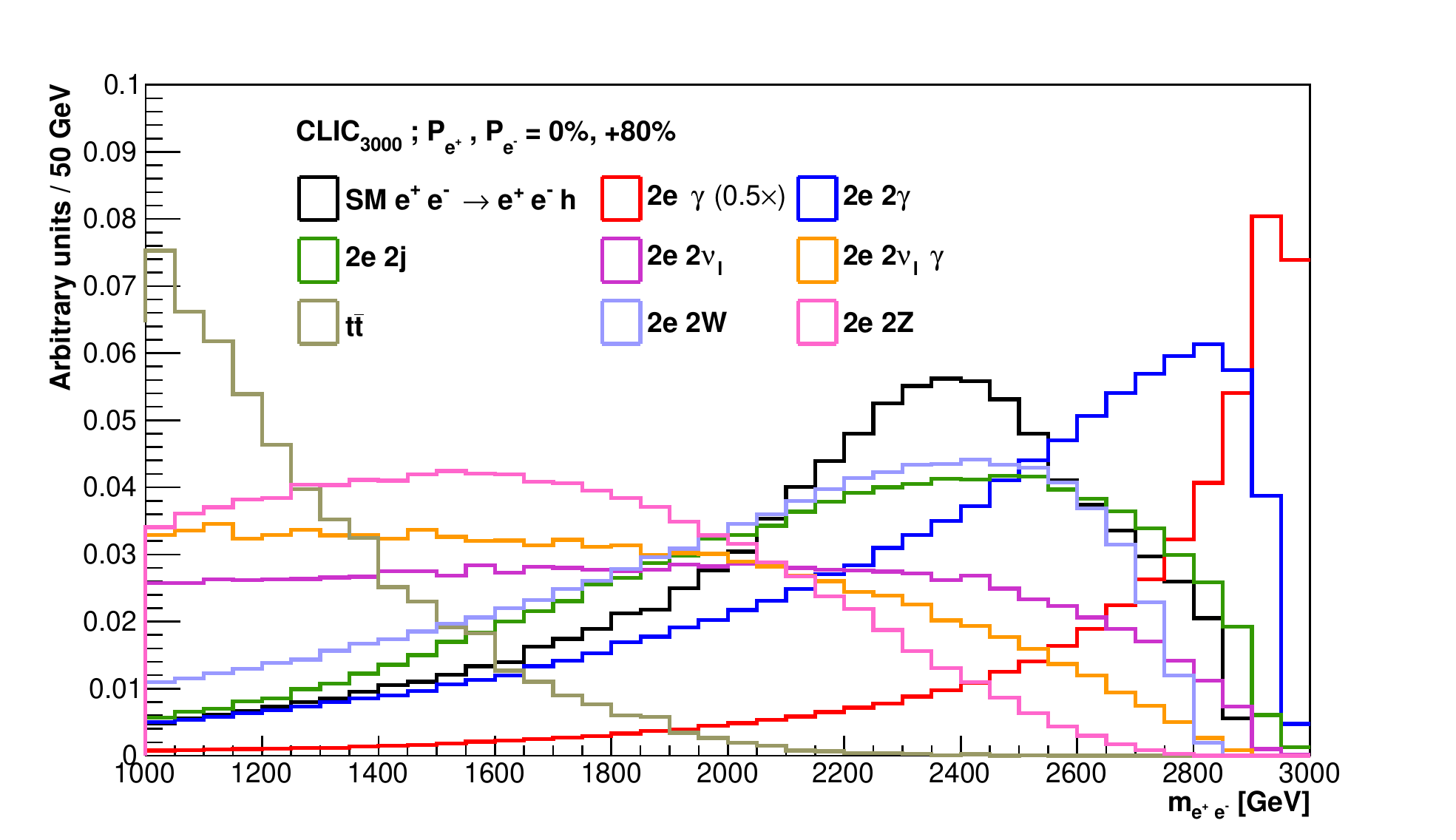}
    \caption{Some plots of the CLIC$_{3000}$ \textit{ZBF} selection with the \textit{right} polarisation for the SM-driven $e^+e^- \to e^+e^-h$ (black), $2e \gamma$ (red), $2e 2\gamma$ (blue), $2e 2j$ (green), $2e 2\nu_{\ell}$ (magenta), $2e 2\nu_{\ell} \gamma$ (orange), $t\overline{t}$ (brown), $2e 2W$ (lilac), and $2e 2Z$ (pink) processes. All histograms are normalised to unity. \textbf{Left:} $\left|\Delta y_{e^+e^-} \right|$ distribution, showing the forward nature of the $e^+e^- \to e^+e^-h$ single-Higgs production process at high energies. \textbf{Right:} $m_{e^+e^-}$ distribution before requiring $m_{e^+e^-} \in [2100.0, 2880.0]$~\gev. The $2e \gamma$ histogram in the right panel is scaled ($0.5\times$) after normalisation for visualisation purposes. Note that, although the \textit{Zh} and \textit{ZBF} diagrams interfere with each other, at high energies the $t$-channel \textit{ZBF} dominates over its $s$-channel counterpart \textit{Zh}.}
    \label{fig:ZBF_selection}
\end{figure}

In practice, we impose a basic set of cuts, beginning with $y^{e^+} \cdot y^{e^-} < 0$ to ensure that the electrons and positrons are in opposite regions of the detector, the fundamental feature of a forward process. Moreover, a large rapidity gap $\left|\Delta y_{e^+e^-}\right|$ is expected, but not observed at the ILC$_{250}$, although it is present at higher energies; at ILC$_{1000}$ (CLIC$_{3000}$) we demand $\left|\Delta y_{e^+e^-}\right| \in [3, 5]$ ($[3.5, 5]$). To further discriminate background events, we require $m_{e^+e^-} \in \{[105, 130], [600, 880], [2100, 2880] \}$~\gev, and to control the effects of ISR, FSR, and beamstrahlung on the shape of the $m_{\text{recoil}}$ distribution, we impose the cut $p_T^{e^+e^-} \in \{[20, 50], [80, 300], [150, 800]\}$~\gev \, at ILC$_{250, 1000}$ and CLIC$_{3000}$, respectively. Finally, from \eq{eq:_RecoilMass} and as done for the \textit{Zh} channel (see Sec. \ref{subsec:_Zhprod}), we apply a hard cut on the recoil system's mass $m_{\text{recoil}} \in [123, 127]$~\gev \, since there is a narrow peak at $m_h \sim 125$~\gev, on top of a continuum background as expected. The $\left|\Delta y_{e^+e^-}\right|$ and $m_{e^+e^-}$ distributions for the CLIC$_{3000}$ with \textit{right} polarisation can be seen in Fig.~\ref{fig:ZBF_selection}, and the impact of the event selection on the cross-sections for the dominant SM processes under consideration is shown in Table~\ref{tab:_CutsZBF} for the three centre-of-mass energies, as well as for the \textit{left} and \textit{right} polarisations. Being a $t$-channel process, the \textit{ZBF} channel dominates over \textit{Zh} at high energies and, as such, the ILC$_{250}$ is not a promising avenue to test the effects of the D6 operators under consideration. As for the \textit{Zh} scenario, we will only focus on ILC$_{1000}$ and CLIC$_{3000}$. As before, the kinematic distributions used to perform the event selection show similar shapes/behaviour in both polarisation settings.

\begin{table}[t]
\centering
\resizebox{\columnwidth}{0.9\height}{
\begin{tabular}{|c|c|c|c|c|c|c|}
\hline
$\sigma_{\sqrt{s}}^{\text{stage}}$ [fb] & SM & $2e\gamma$ & $2e 2\gamma$ & $2e 2\nu_{\ell}$ & $2e 2\nu_{\ell} \gamma$ & $2e 2j$ \\
\hline
\hline
\hline
$\sigma^{\text{in}}_{250}$ & 0.88/0.66 & 47354.7/46966 & 628.53/620.28 & 1348.33/99.46 & 59.79/4.29 & 125.27/115.97 \\
$\sigma^{\text{out}}_{250}$ & 0.26/0.19 & $< 10^{-4}$ & 0.37/0.34 & 1.57/0.13 & 0.11/0.01 & 0.033/0.024 \\
\hline
$\sigma^{\text{in}}_{1000}$ & 14.02/10.54 & 9651.21/9221.23 & 394.58/376.18 & 430.13/59.66 & 36.89/5.04 & 93.29/76.42 \\
$\sigma^{\text{out}}_{1000}$ & 2.52/1.92 & $< 10^{-4}$ & 0.034/0.030 & 0.099/0.016 & 0.0045/0.0017 & 0.024/0.012 \\
\hline
$\sigma^{\text{in}}_{3000}$ & 4.11/3.08 & 1754.91/1631 & 115.66/107.46 & 154.04/29.84 & 17.45/3.56 & 42.19/33.55 \\
$\sigma^{\text{out}}_{3000}$ & 0.22/0.15 & $< 10^{-4}$ & 0.0052/0.0054 & 0.0084/0.0022 & 8.4/4.7 $\left(\times 10^{-4}\right)$ & 0.0033/0.0018 \\
\hline
\end{tabular}}
\caption{Cross-sections, $\sigma$, in fb, for the \textit{ZBF}-like SM-driven signal $e^+e^- \to e^+ e^- h$ and its dominant backgrounds before (\textit{in}) and after (\textit{out}) event selection at $\sqrt{s} = \{250, 1000, 3000\}$~\gev. We consider the \textit{left} and \textit{right} beam polarisations $P_{e^+}$,$P_{e^-}$ for each $\sqrt{s}$, reported as \textit{left/right} (see text for details). For the $t\bar{t}$, $e^+e^-W^+W^-$, and $e^+e^-ZZ$ channels, $\sigma^{\text{in}} \; \textrm{(in fb)} \sim$  1.52/2.38, 15.03/2.69, and 0.04/0.02, respectively at 1 TeV. At 3 TeV, the respective numbers are 0.12/0.21, 29.84/3.88, and 0.08/0.05. For all cases, $\sigma^{\text{out}} < 10^{-4}$ fb. For $\sqrt{s} = 250$ GeV, the kinematic phase space is not open for any of these three channels.}
\label{tab:_CutsZBF}
\end{table}

\section{Projected sensitivities to EFT couplings}
\label{sec:Results}

\begin{figure*}[htp]
\centering
\subfigure[]{\includegraphics[scale=0.55]{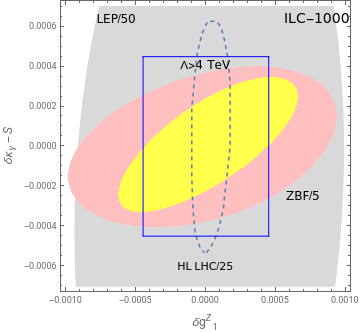}}\quad
\subfigure[]{\includegraphics[scale=0.55]{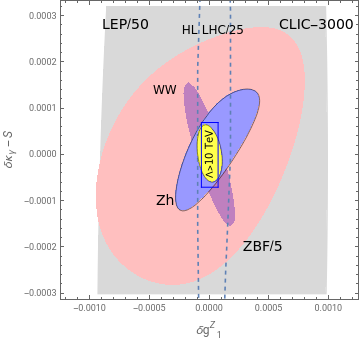}}
\caption{(a)  Projected sensitivities for the case of universal new physics for ILC$_{1000}$ and their comparison with LEP bounds~\cite{LEP} and HL-LHC projections~\cite{Grojean:2018dqj} (b) Projected sensitivities for the case of universal new physics for CLIC$_{3000}$ and their comparison with LEP bounds and HL-LHC projections. We have assumed $W=Y=0$. See the text for more details.}
\label{figuni}
\end{figure*}

\begin{figure*}[htp]
\centering
\subfigure[]{\includegraphics[scale=0.75]{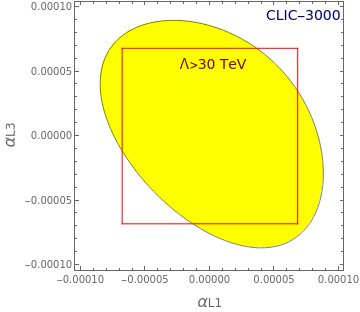}}\quad
\subfigure[]{\includegraphics[scale=0.75]{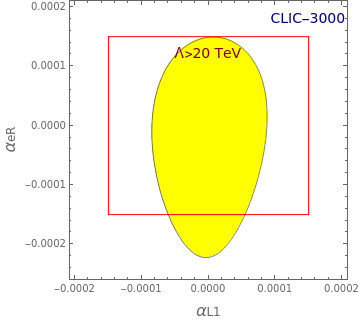}}
\caption{Projected sensitivities for the leptonic high-energy primaries, $\alpha_{e_R}=-c^e_R v^2/{\Lambda^2},~\alpha_{L1}=-c_L^{l,(1)}v^2/{\Lambda^2}~\alpha_{L3}=-{c_L^{l,(3)}v^2}/{\Lambda^2}
$, at CLIC$_{3000}$ in 2-dimensional planes where the third parameter has been marginalised over. See the text for more details }
\label{fig3}
\end{figure*}

\begin{figure}[t]
\centering
\includegraphics[width=0.5\textwidth]{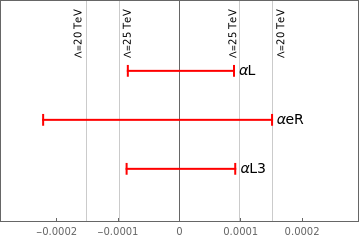}
\caption{Projected sensitivities on the individual leptonic high-energy primaries, $\alpha_{e_R}=-c^e_R v^2/{\Lambda^2},~\alpha_{L1}=-c_L^{l,(1)}v^2/{\Lambda^2}~\alpha_{L3}=-{c_L^{l,(3)}v^2}/{\Lambda^2}
$, at CLIC$_{3000}$ where the other two parameters have been marginalised over. See the text for more details.}
\label{fig5}
\end{figure}

In this section, we present the final sensitivity projections for the EFT couplings. We first generate samples with the EFT couplings turned on and apply the analysis strategy described in the previous section to obtain these. Then, for both the \emph{Zh} and \emph{ZBF} processes, we generate samples for enough points in the EFT parameter space so that it is possible to ascertain all the interference and EFT squared contributions as a function of the Wilson coefficients of the operators in Table~\ref{tab:operators}.

In the following, we will include both the interference term between the SMEFT and SM amplitudes, and the SMEFT squared term. We have checked that, both for the ZBF and the $Zh$ processes, the squared term is between 1$\%$ and 10$\%$ of the interference term if we consider all possible polarisations and center of mass energies. This shows that dimension-eight (D8) effects, which are of the order of the squared term for ${\cal O}(1)$ Wilson coefficients, are indeed negligible. This can also be inferred from the fact that the final scales probed in this study are much larger than the centre-of-mass energy of the process (so that the ratio of D8 to D6 effects, ${s}/{\Lambda^2}\ll 1$).

First, let us consider the \emph{Zh} process. The experimental strategy used in Sec.~\ref{sec:_Collider} to isolate the SM signal with respect to the other backgrounds is sufficient also to isolate the EFT signal. This is because, as discussed in Sec.~\ref{sec:eft}, the high-energy EFT contributions differ from the SM amplitudes only due to the lack of the $Z$-propagator, which implies that the EFT and SM contributions have the same angular dependence (see also Ref~\cite{Banerjee1}). In any case, the dominant background for the EFT signal is by far the SM Higgs-strahlung process. To derive the projected sensitivity for the EFT couplings, we define a $\chi^2$ function as follows,
\begin{eqnarray}
\chi^2_{Zh} = \frac{(N^{exp}_{Zh}- N^{obs}_{Zh})^2}{\sigma_{Zh}^2}\ ,\nonumber
\end{eqnarray}
 where we have taken the SM to be our null hypothesis and $N^{exp}_{Zh}$, denotes the expected number of events in the SM. The number of events observed, $N^{obs}_{Zh}$ will be assumed to be different from the SM due to the presence of EFT couplings. Finally, $\sigma_i$ quantifies the error including both the statistical and systematic uncertainties, 
\bea
\sigma_{Zh}=\sqrt{N^{exp}_{Zh}+(\Delta_{sys} N^{exp}_{Zh})^2}\ ,\nonumber
\eea
where we take the percentage systematic uncertainty to be $\Delta_{sys}=0.03$ following Ref.~\cite{deBlas:2018mhx}. 

As far as the \emph{ZBF} process is concerned, once again, the experimental strategy required to isolate the SM (\emph{ZBF}) Higgs contribution from the other background processes in the previous section would also separate our signal, the EFT contribution. This is again because the dominant high-energy EFT contribution due to the operators in Table~\ref{tab:operators} have an amplitude that is the same as the SM apart from a quadratic growth with respect to the Mandelstam variable $t$ (see also Ref~\cite{Araz:2020zyh}). We will use this quadratic growth with $t$ to distinguish the EFT contribution from the SM. To this end, we will use $p^h_T$, a variable highly correlated to $t$, as the discriminant. To obtain the projected sensitivity for the EFT couplings, we define the following  $\chi^2$ function.
\begin{eqnarray}
\chi^2_{ZBF} = \sum_{i}^{N}\frac{(N^{exp}_i - N^{obs}_i)^2}{\sigma_i^2}\ ,\nonumber
\end{eqnarray}
where we again take the SM as our null hypothesis and, $N^{exp}_i$, the expected number of  events, is taken to be the SM value for the $i$th bin in the $p^h_T$ distribution. The number of events observed in the $i$-th bin, $N^{obs}_i$ would be then assumed to be different from the SM due to the presence of EFT couplings. The total uncertainty for each bin, $\sigma_i$, is given by, 
\bea
\sigma_i=\sqrt{N^{exp}_i +(\Delta_{sys} N^{exp}_i)^2}\ .\nonumber
\eea
 
Finally, for the $e^+e^- \to W^+W^-$ process, the  $\chi^2$ function has been inferred from Ref.~\cite{deBlas:2018mhx} such that the bounds in Table 15 of this paper are reproduced for both the semileptonic and fully hadronic cases. In order to statistically combine both these sub-processes, we then add the $\chi^2$ function for each case to obtain $\chi^2_{WW}$. To obtain our final bounds, we sum over the $\chi^2$ functions for all the different processes, 
 \bea
 \chi^2_{total}=\sum \chi^2_{Zh}+\sum \chi^2_{ZBF}+\sum \chi^2_{WW}
 \eea
where the summation in the terms above is over the different polarisation settings shown for $\sqrt{s}=1,3$~\tev \, in Table~\ref{tab:_SummaryColliders}.

\paragraph{Combination of all channels for universal case}

First, we derive the bounds in the scenario where new physics contributions can be encoded in universal operators of Table~\ref{tab:operators}. We can then present our bounds in a 2-dimensional plane by expressing the Wilson coefficients of the operators in the first column of Table~\ref{tab:operators} as a function of 2 the universal parameters, $(\delta {\kappa}_\gamma-\hat{S})$ and $\delta {g}^Z_1$, where we have ignored the $W$ and $Y$ contributions in \eq{equni}. The latter assumption is reasonable, first of all, because the $W$ and $Y$ parameters are expected to be constrained more strongly in Drell-Yan processes~\cite{deBlas:2018mhx, Torre:2020aiz, Panico:2021vav}. Secondly, there are some well-motivated UV examples where $W$ and $Y$ are small, as discussed in Ref.~\cite{Franceschini:2017xkh}.

We show the results for $\sqrt{s}= 3$~\tev \, ($\sqrt{s}=1$~\tev) in Fig.~\ref{figuni} which includes the bounds from LEP~\cite{LEP} and projected sensitivities for HL-LHC from Ref.~\cite{Grojean:2018dqj}. It can be seen that in both cases, while the \emph{ZBF} produces very strong per-mile level bounds that far surpass the LEP bounds, it is still an order of magnitude smaller than the final bounds, including the process. This is because, while the total cross-section for the \emph{ZBF} process for both $\sqrt{s}=1$~\tev \, and $\sqrt{s}=3$~\tev \, is bigger than the Higgs-strahlung process, the number of events at high $p^h_{T}$ is much smaller; for instance the number of events becomes negligible for $p^h_{T}>300$~\gev \, ($p^h_{T}>800$~\gev) in the $\sqrt{s}= 1$~\tev \, ($\sqrt{s}=3$~\tev) case. In the $\sqrt{s}= 1$~\tev \, case, the final bounds are indistinguishable from the bounds obtained just from the Higgs-strahlung process, which is much more sensitive than the \emph{ZBF} process. For the $\sqrt{s}= 3$~\tev \, case, the bounds from the $e^+e^-\to W^+W^-$ process turn out to be also very important as shown in Fig.~\ref{figuni} (b).

The LEP bounds are weaker by at least two orders of magnitudes with respect to our final bounds and thus have to be shrunk by a factor of 50 to be shown in the same plot \footnote{Note that the LEP bounds were derived assuming that the TGC couplings $\lambda_\gamma=\lambda_Z=0$. The bounds would be even weaker if these two couplings were not assumed to vanish but marginalised over.}. For the universal case, the processes considered here, and the corresponding ones for hadronic colliders with a $pp$ initial state, receive corrections from the same operators shown in the first column of Table~\ref{tab:operators}. This is what allows us to put bounds from HL-LHC, taken from Ref.~\cite{Grojean:2018dqj}, in the same plane. Again our bounds are stronger by one order of magnitude than these HL-LHC projections.

In Fig.~\ref{figuni} we also show a region enclosed by a blue box to show the energy scale, 
\bea
\Lambda \sim \frac{m_W}{\sqrt{\delta {g}^Z_1 \ctw^2}}, ~~\frac{m_W}{\sqrt{\delta {\kappa}_\gamma-\hat{S}}}
\eea
that can be accessed in each case (see \eq{eqsilh} and \eq{equni}). We see that scales as large as 10~\tev \, (4~\tev) can be accessed in the $\sqrt{s}= 3$~\tev \, ($\sqrt{s}=1$~\tev) case. This also shows that our study respects EFT validity considerations. The scales that can be probed here can be compared with the most powerful bounds from LEP on the $\hat{S}$-parameter which translates to the scales around $
\Lambda \sim 1.6-2.5$~\tev~\cite{Giudice:2007fh}.

\paragraph{Combination of all channels for general case}

We now consider the general case where three linearly independent combinations of the leptonic high-energy primaries contribute to the processes considered in this work \footnote{Note that this does not imply that the three high-energy primaries are statistically independent parameters.}. As already discussed in Sec.~\ref{sec:eft}, the three processes are sensitive to different combinations of the Wilson Coefficients of the operators in Table~\ref{tab:operators}. It is clear from \eq{eqprimaries} and \eq{eqprimariesZBF}  that the inclusion of the $W^+W^-$ processes now becomes crucial to leave no flat direction unconstrained because both the Higgs-strahlung and \emph{ZBF} processes are not sensitive to the EFT direction $(\alpha_{L1}-\alpha_{L3})$. In Fig.~\ref{fig3} we provide the 2 dimensional bounds after marginalising over the third parameter. The individual 95 $\%$ CL bounds on each of the three couplings after marginalising over the other two are as follows, 
\bea
\alpha_{L1} &\in&\left[-8.5,8.8\right] \times 10^{-5}\nonumber\\
\alpha_{L3} &\in&\left[-9,9\right]\times 10^{-4}\nonumber\\
\alpha_{e_R} &\in&\left[-2.2,1.5\right] \times 10^{-5}
\label{power}
\eea
We show the same bounds graphically in  Fig.~\ref{fig5}. In both Fig.~\ref{fig3} and Fig.~\ref{fig5} we provide information about the energy-scale, $\lambda\sim v/\sqrt{\alpha_i}$, that can be probed. We see that scales as high as 20-30~\tev \, can be probed.

The Wilson coefficients constrained in Fig.~\ref{fig3} and Fig.~\ref{fig5} can be expressed as a function of pseudo-observables already measured at LEP, as shown in \eq{eq:relation} which we rewrite here for convenience, 
\bea
\alpha_{L1}&=&\frac{\ctw}{g}(\delta g^Z_{e_L}+\delta g^Z_{\nu_L})+\stw^2 \delta g ^Z_1-\ttw^2 \delta \kappa_\gamma\\
\alpha_{L3}&=&\frac{\ctw}{g}(\delta g^Z_{e_L}-\delta g^Z_{\nu_L})+\ctw^2 \delta g ^Z_1\\
\alpha_{e_R}&=&\frac{2 \ctw}{g}\delta g^Z_{e_R}+2 \stw^2 \delta g ^Z_1-2 \ttw^2 \delta \kappa_\gamma.
\eea
We can then use the above relations to compare our bounds with existing LEP bounds or projected HL-LHC sensitivities. The bounds on the right-hand side of the above relations arise mainly from the LEP (or projected HL-LHC) bounds on the TGCs, which are at least 100 (10) times weaker than the bounds in \eq{power}. One can imagine a UV scenario where the EFT contribution to the TGCs vanishes and the bound on the right-hand side above arises from leptonic decays of the $Z$-boson, which were measured very precisely at LEP. These bounds are given by~\cite{Falkowski:2014tna},
\bea
\delta g^Z_{e_L} &\in&\left[-1,9\right] \times 10^{-4}\nonumber\\
 \delta g^Z_{e_R}&\in&\left[-4,2\right] \times 10^{-4}
\eea
We see that the bounds in \eq{power} are more powerful  compared to even the LEP $Z$-pole bounds above.

\section{Conclusions}
\label{sec:Conclusions}

Using the multi-dimensional space of SMEFT has now become the standard way to parametrise indirect effects at LHC and other future colliders~\cite{Buchmuller:1985jz, Giudice:2007fh, Grzadkowski:2010es, Gupta:2011be,Gupta:2012mi,Banerjee:2012xc, Gupta:2012fy, Banerjee:2013apa, Gupta:2013zza, Elias-Miro:2013eta, Contino:2013kra, Falkowski:2014tna, Englert:2014cva, Gupta:2014rxa, Amar:2014fpa, Buschmann:2014sia, Craig:2014una, Ellis:2014dva, Ellis:2014jta, Banerjee:2015bla, Englert:2015hrx, Ghosh:2015gpa, Degrande:2016dqg, Cohen:2016bsd, Ge:2016zro, Contino:2016jqw, Biekotter:2016ecg, deBlas:2016ojx, Denizli:2017pyu, Barklow:2017suo, Brivio:2017vri, Barklow:2017awn, Khanpour:2017cfq, Englert:2017aqb, panico, Franceschini:2017xkh, Banerjee1, Grojean:2018dqj,Biekotter:2018rhp, Goncalves:2018ptp,Gomez-Ambrosio:2018pnl, Freitas:2019hbk, Banerjee:2019pks, Banerjee:2019twi, Biekotter:2020flu, Araz:2020zyh, Ellis:2020unq, Banerjee:2020vtm, Almeida:2021asy, Chatterjee:2021nms}. Most SMEFT studies now aim to include all possible operators contributing to the list of considered processes. While such detailed and systematic studies are perhaps the best way to summarise the results of indirect searches comprehensively, one is often interested in a much simpler question: what is the highest scale that the collider can probe? It is sufficient to include only the most sensitive effects that dominate at high energies to answer this question. In this work, we answer this question in the context of electroweak processes at high-energy lepton colliders like the ILC$_{1000}$ and CLIC$_{3000}$. To this end, we identify the leading EFT effects that grow with energy and perform a `high-energy fit' by including only the corresponding operators.

The processes we include are shown in Table~\ref{tab:process}. Three linear combinations of the operators in Table~\ref{tab:operators}, the so-called leptonic high energy primaries, contribute to these processes at high-energies as shown in \eq{eqprimaries}, \eq{eqsilh}, \eq{eqwarsaw} and \eq{eqprimariesZBF}. These effects are larger than other EFT effects that do not grow with energy by a factor $s/m_Z^2$ which translates to two orders of magnitude for ILC$_{1000}$ and CLIC$_{3000}$~\footnote{Even EFT effects that grow linearly with energy with respect to the leading SM piece will be smaller than these effects by one order of magnitude.}.

Our final sensitivity estimates are shown in Fig.~\ref{figuni}-\ref{fig5} and \eq{power}. These estimates surpass existing LEP bounds by at least two orders of magnitude and projections for HL-LHC by at least an order of magnitude.
 We have also shown in Fig.~\ref{figuni}-\ref{fig5}, the corresponding scales that these colliders can access. We see that the effects studied in this work can probe scales up to tens of~\tev, which corresponds to about $10^{-20}-10^{-19}$~m. This would be the highest energy scale, and the smallest length scale, probed in Higgs/electroweak physics making these lepton colliders the ultimate microscopes to study fundamental physics.

\begin{acknowledgments}

We thank Keith R.K. Ellis, Rikkert Frederix, Valentin Hirschi, Shilpi Jain, Olivier Mattelaer, Rachel C. Rosten, and Zhuoni Qian for helpful exchanges. SB acknowledges the support received from IPPP, Durham, UK, where a part of this work was performed. SB is also grateful for the computing support received from IPPP, Durham. OOV acknowledges the grant received from the Mexican National Council for Science and Technology (CONACYT) – grant number – 460869.
\end{acknowledgments}

\appendix
\section{Contribution of operators to anomalous couplings}
\label{sec:_Section3}

In this appendix we connect the Wilson coefficients of the operators in Table~\ref{tab:operators} to anomalous couplings for both the universal and general case. For the universal case, the relevant Lagragian  is, 
\bea
\Delta {\cal L}_{univ} &=&  -\frac{\hat{T}}{2}\frac{m_Z^2}{2} Z_\mu Z^\mu - \frac{\hat{S}}{4 m_W^2} \frac{ g g' v^2}{2}(W^3_{\mu \nu}B^{\mu \nu})-\frac{W}{2 m_W^2}   ( \partial^\mu  W_{\mu \nu}^3)^2-\frac{Y}{2 m_W^2}  (  \partial^\mu  B_{\mu \nu})^2\nonumber\\
&+& i g \, \delta g_1^Z c_{\theta_W} Z^\mu \left( W^{+\nu} {W}^-_{\mu\nu}  - W^{-\nu} {W}^+_{\mu\nu} \right) + i g \left( \delta\kappa_z c_{\theta_W} {Z}^{\mu\nu} + \delta\kappa_\gamma s_{\theta_W} \hat{A}^{\mu\nu} \right) W^+_\mu W^-_\nu, \nonumber\\
\label{ewpt}
\eea
where at D6 level the following relationship holds,  $\delta \kappa_Z =  \delta g_1^Z - t^2_{\theta_W} \delta \kappa_\gamma$.  Here $\hat{S}$ and $\hat{T}$ are the Peskin-Takeuchi parameters~\cite{Peskin:1991sw},  $W$ and $Y$ are two-other Electroweak Precision observables defined in Ref.~\cite{Barbieri:2004qk} and the anomalous TGCs $\delta g_1^Z$ and $\delta \kappa_\gamma$ were first defined in Ref.~\cite{Hagiwara:1986vm}. The contribution  of the Wilson coefficients of the universal operators in the SILH basis is given by, 
 \bea 
&& \hat{T} =\frac{v^2}{\Lambda^2} c_T \ , \quad 
	\hat{S}=  \frac{m_W^2}{\Lambda^2} \left(c_W + c_B \right) \  \qquad Y  = \frac{m_W^2}{\Lambda^2} c_{2B} \  \qquad
	W= \frac{m_W^2}{\Lambda^2} c_{2W} \   \nonumber\\
	&& \delta g_1^Z =   - \frac{m_W^2}{\Lambda^2} \frac{1}{c_{\theta_W}^2} (c_W+c_{HW})  \  \qquad
	\delta \kappa_\gamma = -\frac{m_W^2}{\Lambda^2} (c_{HW}+c_{HB}),
 \label{eq:ObsCoeffEWPT} 
 \eea
 in the $\{\alpha_{em}, G_F, m_Z\}$ scheme where $c_T$ is the Wilson coefficient of the operator ${\cal O}_T=(H^\dagger \lra{D}_\mu H)^2/2$. From the above equations we can derive the following relationships,
\bea
\ttw^2\left(\delta \kappa_\gamma-\hat{S}-\delta g^Z_1 \ctw^2+Y\right)&=&- \frac{m_W^2}{\Lambda^2}\left(c_B+c_{HB}-c_{2B}\right)\nn\\
\left(\delta g^Z_1 \ctw^2+W\right)&=&-\frac{m_W^2}{\Lambda^2}\left(c_W+c_{HW}-c_{2W}\right)
\eea
which have been used in Sec.~\ref{sec:eft}  and also in Sec.~\ref{sec:Results} to project our results into the plane in Fig.~\ref{figuni}.

We can repeat the same exercise for the general case where, following Ref.~\cite{Gupta:2014rxa}, we now use the Lagrangian, 
\bea
\Delta {\cal L}_6&=&    \delta g^W_{L}\,(W^+_\mu \bar{\nu}^e_L \gamma^\mu e_L+h.c.)+ g^h_{WL}\,\frac{h}{v}(W^+_\mu \bar{\nu}_L \gamma^\mu e_L+h.c.)+ \sum_l \delta g^Z_{l} Z_\mu \bar{l} \gamma^\mu l +\sum_l g^h_{Zl}\,\frac{h}{v}Z_\mu \bar{l} \gamma^\mu l \nn\\&+& i g \, \delta g_1^Z c_{\theta_W} Z^\mu \left( W^{+\nu} \hat{W}^-_{\mu\nu}  - W^{-\nu} \hat{W}^+_{\mu\nu} \right) + i g \left( \delta\kappa_z c_{\theta_W} \hat{Z}^{\mu\nu} + \delta\kappa_\gamma s_{\theta_W} \hat{A}^{\mu\nu} \right) W^+_\mu W^-_\nu\nonumber\\
\label{anam}
\eea
where, for brevity, we have only included the first generation fermions, so that $l=e_L, e_R, \nu^e_L$ and $L$ is the first-generation lepton doublet. 
 
The operators of the  Warsaw basis~\cite{Grzadkowski:2010es} in the right panel of Table~\ref{tab:operators},  give the following contributions to these vertices, 
 \bea
 \label{wilson}
 \delta g^W_{l}&=& \frac{g}{\sqrt{2}}\frac{v^2}{\Lambda^2} c_L^{l,(3)}+\frac{\delta m^2_Z}{m^2_Z}\frac{\sqrt{2}g\ctw^2}{4\stw^2}\nn\\
   g^h_{Wf}&=& \sqrt{2} g\frac{v^2}{\Lambda^2} c_L^{l,(3)}\nn\\
   \delta g^Z_{f}&=&-\frac{g Y_f \stw}{\ctw^2}\frac{v^2}{\Lambda^2} c_{WB} -\frac{g}{\ctw}\frac{v^2}{\Lambda^2}(|T_3^f|c_L^{l,(1)}-T_3^f c_L^{l,(3)}+(1/2-|T_3^f|)c^e_{R})\nn\\&+&\frac{\delta m^2_Z}{m^2_Z}\frac{g}{2\ctw\stw^2}(T_3 \ctw^2+Y_f \stw^2)\nn\\
  g^h_{Zf}&=&- \frac{2 g}{\ctw}\frac{v^2}{\Lambda^2}(|T_3^f|c_L^{l,(1)}-T_3^f c_L^{l,(3)}+(1/2-|T_3^f|)c^e_{R})\nn\\
  \delta g^Z_{1}&=&\frac{1}{2 \stw^2}\frac{\delta m_Z^2}{m_Z^2}\nn\\
\delta\kappa_{\gamma}&=&\frac{1}{\ttw} \frac{v^2}{\Lambda^2} c_{WB}\,,
 \eea
where we have now used $(m_W, m_Z,\alpha_{em})$ as our input parameters following Ref.~\cite{Gupta:2014rxa}.  In the above equations, the term,
\bea
\frac{\delta m^2_Z}{m^2_Z}= \frac{v^2}{\Lambda^2}(2 \ttw c_{WB}+\frac{c_{HD}}{2}),
\eea
corresponds to  the shift in the input parameter, $m_Z$,  due to the  operators ${\cal O}_{WB}$ and ${\cal O}_{HD}$ defined in Ref.~\cite{Banerjee:2019twi}. Using the above equations, one can derive the following relationships, 
\bea
c_L^{l,(1)} \frac{v^2}{\Lambda^2}&=&-\frac{\ctw}{g}(\delta g^Z_{e_L}+\delta g^Z_{\nu_L})-\stw^2 \delta g ^Z_1+\ttw^2 \delta \kappa_\gamma\\
c_L^{l,(3)} \frac{v^2}{\Lambda^2}&=&-\frac{\ctw}{g}(\delta g^Z_{e_L}-\delta g^Z_{\nu_L})-\ctw^2 \delta g ^Z_1\\
c^e_{R} \frac{v^2}{\Lambda^2}&=&-\frac{2 \ctw}{g}\delta g^Z_{e_R}-2 \stw^2 \delta g ^Z_1+2 \ttw^2 \delta \kappa_\gamma
\label{eq:relationC}
\eea
that have been used in Sec.~\ref{sec:eft} and Sec.~\ref{sec:Results}.

\bibliography{Biblio}

\end{document}